\documentclass[aps,pra,amsmath,amssymb,twocolumn,superscriptaddress,showpacs]{revtex4-1}
\usepackage{amsmath,amssymb}
\usepackage{graphicx}	

\def\be{\begin{eqnarray}}   
\def\ee{\end{eqnarray}}

\begin{document}

\author{S.~A.~Sato}
\email{ssato@ccs.tsukuba.ac.jp}
\affiliation 
{Center for Computational Sciences, University of Tsukuba, Tsukuba 305-8577, Japan}
\affiliation 
{Max Planck Institute for the Structure and Dynamics of Matter, Luruper Chaussee 149, 22761 Hamburg, Germany}

\author{U.~De~Giovannini}
\affiliation 
{Max Planck Institute for the Structure and Dynamics of Matter, Luruper Chaussee 149, 22761 Hamburg, Germany}

\author{S.~Aeschlimann}
\affiliation 
{Max Planck Institute for the Structure and Dynamics of Matter, Luruper Chaussee 149, 22761 Hamburg, Germany}

\author{I.~Gierz}
\affiliation 
{Max Planck Institute for the Structure and Dynamics of Matter, Luruper Chaussee 149, 22761 Hamburg, Germany}

\author{H.~H\"ubener}
\affiliation 
{Max Planck Institute for the Structure and Dynamics of Matter, Luruper Chaussee 149, 22761 Hamburg, Germany}

\author{A.~Rubio}
\email{angel.rubio@mpsd.mpg.de}
\affiliation 
{Max Planck Institute for the Structure and Dynamics of Matter, Luruper Chaussee 149, 22761 Hamburg, Germany}
\affiliation 
{Center for Computational Quantum Physics (CCQ), Flatiron Institute, 162 Fifth Avenue, New York, NY
10010, USA}

\title{Floquet states in dissipative open quantum systems
}

\begin{abstract}
We theoretically investigate basic properties of nonequilibrium
steady states of periodically-driven open quantum systems based on 
the full solution of the Maxwell-Bloch equation.
In a resonantly driving condition, we find that the transverse relaxation, 
also known as decoherence, significantly destructs the formation of Floquet states 
while the longitudinal relaxation does not directly affect it.
Furthermore, by evaluating the quasienergy spectrum of the nonequilibrium steady states,
we demonstrate that the Rabi splitting can be observed as long as 
the decoherence time is as short as one third of the Rabi-cycle.
Moreover, we find that Floquet states can be formed even under significant dissipation
when the decoherence time is substantially shorter than the cycle of driving,
once the driving field strength becomes strong enough.
In an off-resonant condition, we demonstrate that the Floquet states can be 
realized even in weak field regimes because the system is not excited and 
the decoherence mechanism is not activated. Once the field strength becomes strong enough, 
the system can be excited by nonlinear processes and the decoherence process becomes active. 
As a result, the Floquet states are significantly disturbed by the environment
even in the off-resonant condition.
Thus, we show here that the suppression of heating is a key condition for the realization of Floquet
states in both on and off-resonant conditions not only because it prevents material damage but also because it contributes to preserving coherence.
\end{abstract}

\maketitle

\section{Introduction \label{sec:intro}}

Nonlinear interaction of strong light with matter is an important subject
from both fundamental and technological points of view and has been intensively investigated
for a long time \cite{PhysRevLett.7.118,butcher1990elements,2008v,MOUROU2012720,Flick3026,Ruggenthaler2018}.
Laser fields can directly couple with electrons in matter and induce
nonequilibrium electron dynamics. Thus, strong laser fields can be
employed to control the electronic properties and functionalities of materials
\cite{RevModPhys.81.163,Krausz2014,Basov2017}.
In the ultrafast regime, light-induced electron dynamics in solids within sub-femtosecond
time-scale have been intensively investigated towards petahertz electronics
\cite{Schiffrin2012,Schultze2012,Schultze1348,Lucchini916,Mashiko2016,Schlaepfer2018,2018arXiv181207420S,2018arXiv181100801V}.
In addition, strong light may also couple with phonons in solids and renormalize 
electron-phonon coupling, triggering light-induced superconductivity
\cite{Fausti189,Mitrano2016}.
In these light-induced phenomena, target systems are actively driven by light.
Therefore, substantial energy transfer from light to matter is expected \cite{Sommer2016}.
In most cases, the systems of interest are not isolated but are coupled to their surrounding environment.
Thus, a part of the transferred energy is dissipated to the environment.
Moreover, through the interaction with the environment, \textit{coherence} of 
light-induced dynamics can be lost.
Therefore, proper understanding of light-driven dynamics in a dissipative environment
is indispensable towards realization of optical-control of realistic systems.

Nonequilibrium dynamics of periodically-driven \textit{non-dissipative}
systems has been intensively investigated with Floquet theory, and
various interesting properties of the driven systems have been discussed
such as the emergence of new topological states \cite{PhysRevB.79.081406},
the dynamical localization \cite{PhysRevB.34.3625,grossmann1992localization},
among others \cite{RevModPhys.89.011004,Shin2018,2019reviewGiovannini,Claassen2019}.
Theoretical investigation of periodically-driven systems
has been further extended to \textit{dissipative} systems, or namely, open quantum systems
\cite{PhysRevE.55.300,GRIFONI1998229,PhysRevLett.117.250401,Hartmann_2017},
and the effects of the dissipative environment have been discussed in various aspects
such as realization of Floquet-Gibbs state
\cite{kohn2001periodic,PhysRevE.91.030101,Shirai_2016,PhysRevLett.123.120602},
the asymptotic states
under the memory-effect \cite{PhysRevA.96.042103,PhysRevE.98.022111},
and topological properties \cite{PhysRevB.90.195429,PhysRevB.91.155422}.
Furthermore, experimental studies have been conducted for strongly driven quantum systems
\cite{PhysRevLett.98.257003,PhysRevLett.115.133601,Magazzu2018}, and 
the Floquet-Bloch states have been experimentally
observed as the Rabi-splitting in the angular-resolved
photo-electron spectroscopy (ARPES) in extended systems \cite{Wang453}.
Recently, the light-induced anomalous Hall effect in graphene
has been experimentally observed \cite{2018arXiv181103522M}, 
originating from population effects
on top of the realization of Floquet states subjected to substantial dissipation
\cite{PhysRevB.99.214302,2019arXiv190512981S}.

Optical-control of materials based on Floquet engineering has been attracting great interest
ranging from light-induced topological phase transitions \cite{PhysRevB.79.081406,Hubener:2017ht} to
optical control of chiral superconductors \cite{Claassen2019}.
However, the realization of Floquet states and the control of their population are highly 
nontrivial tasks in a dissipative environment \cite{PhysRevE.79.051129,PhysRevX.5.041050}.
In this work, we theoretically investigate basic properties of periodically-driven
open-quantum systems with Maxwell-Bloch equation \cite{1072212,meier2007coherent},
which may be the simplest
model for driven open-quantum systems, and provide an insight into the realization
of Floquet states in such systems, addressing the following open questions:
Which kind of relaxation mechanism affects the formation of Floquet states, and which does not?
How long is coherence required to persist to realize Floquet states?
Can Floquet states be formed even under a significant dissipation, e.g.
when the relaxation time is shorter than a driving period?

The paper is organized as follows: In Sec.~\ref{sec:method} we first describe
the Maxwell-Bloch equation and several equivalent descriptions of
open quantum systems. Then, we introduce basic quantities of nonequilibrium
steady states, Floquet fidelity and quasienergy spectrum, which
will be investigated in the following sections.
In Sec.~\ref{sec:results} we investigate the properties of the nonequilibrium
steady state in both resonant and off-resonant conditions.
Finally, our findings are summarized in Sec.~\ref{sec:summary}.

\section{Method \label{sec:method}}

Here, we describe theoretical methods to investigate basic properties of 
nonequilibrium steady states of open quantum systems under a periodic driving. 
First, we introduce the Maxwell-Bloch equation, where time-evolution of a density matrix of 
a two-level system under a periodic driving is described with
relaxation time approximation. Then, we revisit equivalent descriptions
of the open-quantum system with the Lindblad equation and the stochastic
Schr\"odinger equation.
Finally, we introduce two quantities to study nonequilibrium steady states;
Quasienergy spectrum of driven open-quantum systems and Floquet fidelity
\cite{PhysRevB.99.214302,2019arXiv190512981S}.

\subsection{Equation of motion of driven open-quantum systems \label{subsec:eom}}

In order to get a microscopic understanding of the nonequilibrium dynamics 
of open quantum systems,
we consider a two-level driven system in dissipative environment based on
the Maxwell-Bloch equation. For detailed analysis of the nonequilibrium
system, we further revisit equivalent descriptions in different forms.

\subsubsection{Maxwell-Bloch equation}

We first revisit the description of nonequilibrium dynamics of a two-level
system based on Maxwell-Bloch equation \cite{1072212},
which can be seen as the simplest form of the semiconductor Bloch equation
and has been used to develop microscopic insight into the driven dynamics
of dissipative systems \cite{meier2007coherent}.
Note that the semiconductor Bloch equation with the simple relaxation time approximation
\cite{PhysRevLett.73.902} has been widely employed in studies on various phenomena
of nonlinear light-matter interactions such as the attosecond electron dynamics
\cite{Moulet1134}, the high-order harmonic generation from solids
\cite{Luu2015,Yoshikawa736} and the light-induced anomalous Hall effect
\cite{PhysRevB.99.214302}. Furthermore, it has been demonstrated that
the simple approximation already performs excellently when compared to the experimental results.
Based on this fact, here we employ the simplest relaxation time approximation
in order to clarify the primary role of the dissipation in driven quantum systems.

The time propagation of
the system is described by the following quantum master equation
\be
\frac{d}{dt}\rho(t) = \frac{\left [ H(t),\rho(t) \right ]}{i\hbar} 
+ \hat D \left [\rho (t)\right ],
\label{eq:master}
\ee
where $\rho (t)$ is the density matrix of the two-level system,
$H(t)$ is the Hamiltonian, and $\hat D \left [\rho (t)\right ]$ is the relaxation
operator.
The Hamiltonian of the two-level system is given by
\be
H(t) = \frac{\Delta}{2}\sigma_z + F_0 \sin (\Omega t) \sigma_x,
\ee
where $\sigma_i$ are Pauli matrices, $\Delta$ is the gap of the two-level system,
$F_0$ is the amplitude of an external field, and $\Omega$ is its frequency.
Furthermore, the dissipation operator is constructed
with the relaxation time approximation, where the relaxation is simply
treated as simpe exponential decays \cite{PhysRevLett.73.902}, as
\be
\hat D \left [\rho (t)\right ] = 
\left(
    \begin{array}{cc}
      -\frac{\rho_{ee}(t)}{T_1} & -\frac{\rho_{eg}(t)}{T_2} \\
      -\frac{\rho_{ge}(t)}{T_2} & -\frac{\rho_{gg}(t)-1}{T_1}
    \end{array}
  \right),
\label{eq:relaxation_time}
\ee
where $\rho_{ij}(t)$ is a matrix element of the density matrix,
$\langle i|\rho(t)|j\rangle$, with the eigenbasis of the subsystem Hamiltonian,
$\Delta \sigma_z/2$; $|g\rangle = (0,1)^{\dagger}$ corresponds to the ground state,
while $|e\rangle = (1,0)^{\dagger}$ corresponds to the excited state.
While the longitudinal relaxation time $T_1$ corresponds to the population decay constant
from the excited state $|e\rangle$ to the ground state $|g\rangle$,
the transverse relaxation time $T_2$ corresponds to the decay constant
for \textit{coherence}. Thus, in this work, we shall call $T_2$
\textit{decoherence time}.

\subsubsection{Lindblad equation}

To smoothly connect the Maxwell-Bloch equation to the equivalent
stochastic Schr\"odinger equation approach \cite{breuer2002theory,2019arXiv190200967L},
which will be used to
introduce the quasienergy spectrum of the driven system later,
here we revisit the relation between the Maxwell-Bloch equation
and the Lindblad equation~\cite{breuer2002theory,2019arXiv190200967L}.
The quantum master equation~(\ref{eq:master})
with the relaxation time approximation~(\ref{eq:relaxation_time})
can be equivalently expressed in the following Lindblad form
\be
\frac{d}{dt}\rho(t) &=& \frac{\left [ H(t),\rho(t) \right ]}{i\hbar} \nonumber \\
&& + \sum^2_{\alpha=1}\gamma_{\alpha}\left (
L_{\alpha}\rho(t)L^{\dagger}_{\alpha}
-\frac{1}{2} \left \{ L^{\dagger}_{\alpha}L_{\alpha}, \rho(t)
\right \}
\right )
\label{eq:lindblad}
\ee
by employing the two Lindblad operators,
\be
L_{1} &=& \sigma_x - i \sigma_y,  \label{eq:L1-operator}\\
L_{2} &=& -\sigma_z. \label{eq:L2-operator}
\ee
Here, the anticommutator is defined as $\{A,B\}=AB+BA$.
The scattering rate~$\gamma_{\alpha}$ in the Lindblad equation~(\ref{eq:lindblad})
and the relaxation time~$T_i$ in the Maxwell-Bloch equation~(\ref{eq:master})
have the following relations,
\be
\frac{1}{T_1} &=& \gamma_1, \\
\frac{1}{T_2} &=& \frac{\gamma_1}{2}+2\gamma_2.
\ee

Note that the Lindblad master equation~(\ref{eq:lindblad}) is invariant under the arbitrary phase multiplication, $L_i\rightarrow e^{i\theta}L_i$.
Nevertheless, we explicitly choose the form of the Lindblad operator, especially for, $L_2$
so that the environment scatters only the excited state $|e\rangle$ but the ground state $|g\rangle$ is not scattered
in the following Stochastic approach.
Preventing scattering of the ground-state also avoids unphysical modification of the quasienergy spectrum of the undriven system that would occur otherwise.

\subsubsection{Stochastic Schr\"odinger equation approach \label{subsubsec:stochastic-wf}}
The Lindblad equation~(\ref{eq:lindblad}) for the density matrix propagation
can be equivalently described 
by a stochastic approach based on non-Hermitian Schr\"odinger equation
for the wavefunction propagation \cite{breuer2002theory,2019arXiv190200967L}.
Therefore, the Maxwell-Bloch equation can be equivalently evaluated
with the stochastic Schr\"odinger equation approach.

For later convenience, we revisit this stochastic trajectory approach
\cite{breuer2002theory,2019arXiv190200967L}.
To introduce the wavefunction propagator, we rewrite the Lindblad
equation~(\ref{eq:lindblad}) as
\be
\frac{d}{dt}\rho(t) = \frac{\left [ H_c(t),\rho(t) \right ]}{i\hbar} 
 + \sum^2_{\alpha=1}\gamma_{\alpha}
L_{\alpha}\rho(t)L^{\dagger}_{\alpha}
\label{eq:lindblad-conditional}
\ee
with the conditional Hamiltonian defined as
\be
H_c(t) \equiv H(t)  
-\frac{i\hbar}{2}\sum^2_{\alpha=1}\gamma_{\alpha}L^{\dagger}_{\alpha}L_{\alpha}.
\ee

The time-evolution of the density matrix $\rho(t)$ obeying the Lindblad 
equation~(\ref{eq:lindblad}) can be obtained by propagating the non-Hermitian Schr\"odinger equation,
$i\hbar d/dt |\tilde \psi(t)\rangle = H_c(t)| \tilde \psi(t) \rangle$,
with stochastic quantum jumps that occur at a given time step.
The probability of the stochastic quantum jump is evaluated from the norm
of the wavefunction, $\langle \tilde \psi(\tau)| \tilde \psi(\tau) \rangle$.
In practical calculations, we employ the the following algorithm.
For simplicity, here we assume that the initial density matrix $\rho(0)$ is 
a pure state: $\rho(0)= |\tilde \psi(0)\rangle \langle \tilde \psi(0)|$.

\begin{description}
\item[(i)] Set the initial wavefunction to the initial pure state $|\tilde \psi(0)\rangle$.

\item[(ii)] Propagate the wavefunction $|\tilde \psi(t)\rangle$ by
the non-Hermitian conditional Hamiltonian $H_c(t)$;
\be
i\hbar \frac{d}{dt}|\tilde \psi(t)\rangle = H_c(t)| \tilde \psi(t) \rangle.
\ee

\item[(iii)] Perform quantum jump at $t_j + \tau$ after time $\tau$
since the last jump at $t_j$ with probability 
$p=\langle \tilde \psi(t_j+\tau)| \tilde \psi(t_j+\tau) \rangle$;
\be
|\tilde \psi(t_j + \tau) \rangle \mapsto \frac{L_{\alpha} |\tilde \psi(t_j + \tau) \rangle}
{\sqrt{\langle \tilde \psi(t_j+\tau)| \tilde \psi(t_j+\tau) \rangle}},
\ee
where the index $\alpha$ is chosen with probability
\be
p_{\alpha}=\frac{\gamma_{\alpha}
\langle \tilde \psi(t_j + \tau)|L^{\dagger}_{\alpha}L_{\alpha} |\tilde \psi(t_j + \tau) \rangle}
{\sum_{\beta} \gamma_{\beta} 
\langle \tilde \psi(t_j + \tau)|L^{\dagger}_{\beta} L_{\beta} |\tilde \psi(t_j + \tau) \rangle}.
\ee

\item[(iv)] Repeat the steps~(ii) and (iii) until the simulation time $t$
reaches the final time of the simulation $t_f$.

\item[(v)] Repeat the above stochastic procedures and evaluate the density matrix
as the statistical average from the stochastic trajectories as
\be
\rho(t) \approx \left < \frac{|\tilde \psi(t)\rangle\langle \tilde \psi(t)|}
{\langle \tilde \psi(t)| \tilde \psi (t)\rangle}
\right >_{average}.
\label{eq:rho-wf-stochastic}
\ee
In the limit of large number of trajectories, 
it can be demonstrated \cite{breuer2002theory,2019arXiv190200967L} that
the statistical average converges
to the solution of the Lindblad equation~(\ref{eq:lindblad}).

\end{description}

Note that the expectation value of an operator $\hat A$ can be evaluated
as the statistical average of the corresponding expectation value of the 
stochastic wavefunctions as
\be
\langle A\rangle &=& \mathrm{Tr}\left [\hat A \rho(t) \right ]
\approx \mathrm{Tr}\left [\hat A  
\left < \frac{|\tilde \psi(t)\rangle\langle \tilde \psi(t)|}
{\langle \tilde \psi(t)| \tilde \psi (t)\rangle}
\right >_{average}
\right ] \nonumber \\
&=&\left < \frac{\langle \tilde \psi(t)| \hat A |\tilde \psi(t)\rangle}
{\langle \tilde \psi(t)| \tilde \psi (t)\rangle}
\right >_{average}.
\label{eq:observable-stochastic}
\ee

\subsection{Quasi-energy spectrum of driven open-quantum systems \label{subsec:pes}}

Here, we introduce a computational scheme to evaluate a quasienergy spectrum 
of driven systems. The scheme is inspired by what is done in the modeling of
photoelectron spectroscopy~(PES), 
which is widely used to investigate
equilibrium quasiparticle energy spectra as well as those of driven systems
\cite{pes_cardona_1,RevModPhys.75.473,Wang453,2019reviewGiovannini}.

First, we describe the method to compute the quasienergy spectrum
of closed driven systems. Later, we will extend it to open quantum systems
based on the above stochastic Schr\"odinger equation approach.
Here, we consider a two level system described
by the following Schr\"odinger equation,
\be
i\hbar\frac{d}{dt}|\psi(t)\rangle
&=&H(t)|\psi(t)\rangle \nonumber \\
&=&
\left [
\frac{\Delta}{2}\sigma_z + F(t)\sigma_x
\right ]|\psi(t)\rangle,
\label{eq:tdse}
\ee
where $F(t)$ is a time-dependent external field.
In order to investigate the quasienergy spectrum of the driven system,
we introduce \textit{theoretical detector states}, $|\epsilon\rangle$,
associated with each detected energy $\epsilon$.
By embedding these theoretical detector states into the original Hilbert space of 
the two-level system, we reconstruct the Hamiltonian of the full system as
\be
H^{E}(t) = && H(t) + \int^{\infty}_{-\infty} d\epsilon 
|\epsilon\rangle \epsilon \langle \epsilon | \nonumber \\
&+& v(t)\int^{\infty}_{-\infty} d\epsilon 
\bigg (
 |\epsilon\rangle\langle i| + |i \rangle \langle \epsilon| 
\bigg ),
\ee
where the first term is the original Hamiltonian of the system that is being probed $H(t)$,
the second term is the Hamiltonian of the embedded detector states $|\epsilon\rangle$,
and the last term is the interaction between the system of interest
and the theoretical detector states via a probe field $v(t)$.
In the interaction Hamiltonian, $|i\rangle$ denotes a state of 
the original system that we would like to probe, and we set it to the ground state $|i\rangle=|g\rangle$.
Furthermore, we employ the following form for the probe field,
\be
v(t) = f(t) \sin \left (\omega t \right ),
\ee
where $\omega$ is the driving frequency of the probe perturbation, and $f(t)$ is 
an envelope function.
Assuming the probe field is weak enough, 
the solution of the Schr\"odinger equation of the full system,
$i\hbar d/dt |\psi^E (t)\rangle = H^E(t) |\psi^E (t)\rangle$
can be approximated by
\be
|\psi^E(t)\rangle \approx |\psi(t)\rangle + \int^{\infty}_{-\infty} d\epsilon \,c(\epsilon,t) 
e^{-i\frac{\epsilon}{\hbar} t}|\epsilon\rangle,
\ee
where $|\psi(t)\rangle$ is the solution of the Schr\"odinger
equation of the original system, Eq.~(\ref{eq:tdse}),
and $c(\epsilon,t)$ is an expansion coefficient of the theoretical detector state
$|\epsilon \rangle$.
Employing the rotating wave approximation,
the equation of motion for the coefficient $c(\epsilon, t)$
can be approximated as
\be
\dot c(\epsilon, t) = \frac{f(t)}{2}\langle i|\psi(t)\rangle 
e^{i\frac{\epsilon-\omega }{\hbar}t}.
\ee
Thus, the population of the detector state $|\epsilon\rangle$ after
the probe perturbation $v(t)$ can be evaluated as
\be
n^{pop}(\epsilon):= \left |c(\epsilon, t=\infty) \right|^2 =
\left |
\int^{\infty}_{-\infty} dt \frac{f(t)}{2}\langle i|\psi(t)\rangle 
e^{i\frac{\epsilon-\omega }{\hbar}t}
\right |^2. \nonumber \\
\label{eq:pop-pes-closed}
\ee
The population distribution at the detector
reflects the quasienergy structure of the closed quantum system described
by the Hamiltonian $H(t)$
as the conventional photoelectron spectroscopy does
\cite{pes_cardona_1,RevModPhys.75.473,Wang453,2019reviewGiovannini}.

We further extend this numerical PES scheme to open quantum systems.
Inspired by a fact that the expectation value of an observable can be evaluated
as the ensemble average of stochastic trajectories
with Eq.~(\ref{eq:observable-stochastic}), we evaluate the quasienergy spectrum
of open quantum systems as the ensemble average of the spectrum of each trajectory.
In practice, the population of the detector states in the case of open quantum systems 
is computed as the statistical average of stochastic trajectories,
\be
n^{pop}(\epsilon)=
\left <
\left |
\int^{\infty}_{-\infty} dt \frac{f(t)}{2}
\frac{\langle i|\tilde \psi(t)\rangle }
{\sqrt{\langle \tilde \psi(t)|\tilde \psi(t)\rangle}}
e^{i\frac{\epsilon-\omega }{\hbar}t}
\right |^2 \right >_{average}. \nonumber \\
\label{eq:pop-pes-open}
\ee

Assuming the quasienergy structure is mapped to the population distribution of the detector
states, $n^{pop}(\epsilon)$, by a single photon absorption with the energy of 
$\hbar \omega$, the quasienergy spectrum $A^{Q}(E_Q)$ as a function of energy $E_Q$
can be evaluated as $A^{Q}(E_q)\sim n^{pop}(E_Q+\hbar \omega)$.

\subsection{Floquet fidelity \label{subsec:floquet-fidelity}}

Here, we introduce a measure to quantify the similarity of
nonequilibrium steady states under periodic driving fields and 
the corresponding Floquet states.
We shall call it \textit{Floquet fidelity} \cite{PhysRevB.99.214302,2019arXiv190512981S}.

Floquet states $|\psi_{F,a}(t)\rangle$ 
are defined as solutions of the time-dependent Schr\"odinger equation with
a time-periodic Hamiltonian, $id/dt|\psi_{F,a}(t)\rangle=H(t)|\psi_{F,a}(t)\rangle$,
with the following form:
\be
|\psi_{F,a}(t)\rangle=e^{-i\frac{\epsilon_{F,a}}{\hbar}t}|u_{F,a}(t)\rangle,
\ee
where $|u_{F,a}(t)\rangle$ has the same time-period as the Hamiltonian, $H(t)$,
and $\epsilon_{F,a}$ is the Floquet quasienergy.
As Floquet states are defined as the solutions of
the time-dependent Schr\"odinger equation,
they are not necessarily solutions of the quantum master 
equation (\ref{eq:master}). Nevertheless, nonequilibrium steady states of
open quantum systems may show some signatures based on the corresponding Floquet states 
under certain conditions \cite{kohn2001periodic,PhysRevE.91.030101,Shirai_2016}. 

To introduce the Floquet fidelity, we first consider the eigenvalue decomposition
of the density matrix in a nonequilibrium steady state as,
\be
\rho(t)=\sum_a \lambda_a(t) |NO_a(t)\rangle\langle NO_a(t)|,
\ee
where $\lambda_a(t)$ is an eigenvalue and $|NO_a(t)\rangle$ is
the corresponding eigenvector.
Since the density matrix of the nonequilibrium steady
state has the time-periodicity of the Hamiltonian, $\rho(t)=\rho(t+2\pi/\Omega)$,
the eigenvalues and the eigenvectors may have the same periodicity,
$\lambda_a(t)=\lambda_a(t+2\pi/\Omega)$ and 
$|NO_a(t)\rangle=|NO_a(t+2\pi/\Omega)\rangle$.
These eigenvectors of the one-body reduced density matrix are known as
\textit{natural orbitals} \cite{PhysRev.97.1474}, and
the eigenvalues can be interpreted as their occupations.
By construction of the natural orbitals, 
the expectation value of an observable $\hat A$ can be
evaluated as the sum of the expectation value of each natural orbital with the occupation
weight as
\be
\langle A \rangle &=& \mathrm{Tr}\left \{ \hat A \rho(t) \right \} \nonumber \\
&=&\sum_a \lambda_a(t) \langle NO_a(t)|\hat A|NO_a(t)\rangle.
\ee
Therefore, the natural orbitals can be seen as very accurate representative
single-particle states
of the system. Based on this fact, we quantify the similarity of the nonequilibrium 
steady state and the Floquet states by the similarity of the corresponding
natural orbitals and the Floquet states.

In practice, to define the similarity, we first introduce a Floquet fidelity matrix $F$
\cite{PhysRevB.99.214302}
whose matrix elements $F_{ij}$ are defined as the cycle average of 
the squared overlap of the $i$-th natural orbital and the $j$-th Floquet state as
\be
F_{ij} = \frac{1}{T} \int^T_0 dt \left |\langle NO_i(t)|\psi_{F,j}(t)\rangle
\right |^2,
\ee
where $T$ is the time-period of the Hamiltonian, $T=2\pi/\Omega$.
Then, the Floquet fidelity $S_F$ is defined as the absolute value of the determinant
of the Floquet fidelity matrix, $S_F=|\mathrm{det}F|$.
The Floquet fidelity takes the maximum value of one only if 
all the natural orbitals have identical Floquet states.
Therefore, if the Floquet fidelity is one, the Floquet states diagonalize the density matrix.
In general, $0\le S_F \le 1$.

\section{Results \label{sec:results}}

\subsection{Resonant driving \label{subsec:resonance}}

We first investigate the nonequilibrium steady state of the two-level system
under the periodic driving in the resonant condition, $\hbar \Omega=\Delta$.
To realize the nonequilibrium steady state, we perform sufficiently long
real-time propagation by solving the Maxwell-Bloch
equation~(\ref{eq:master}). Here, the initial condition is set to the ground state,
$\rho(t=0)=|g\rangle \langle g|$.

\begin{figure}[htbp]
  \includegraphics[width=0.95\columnwidth]{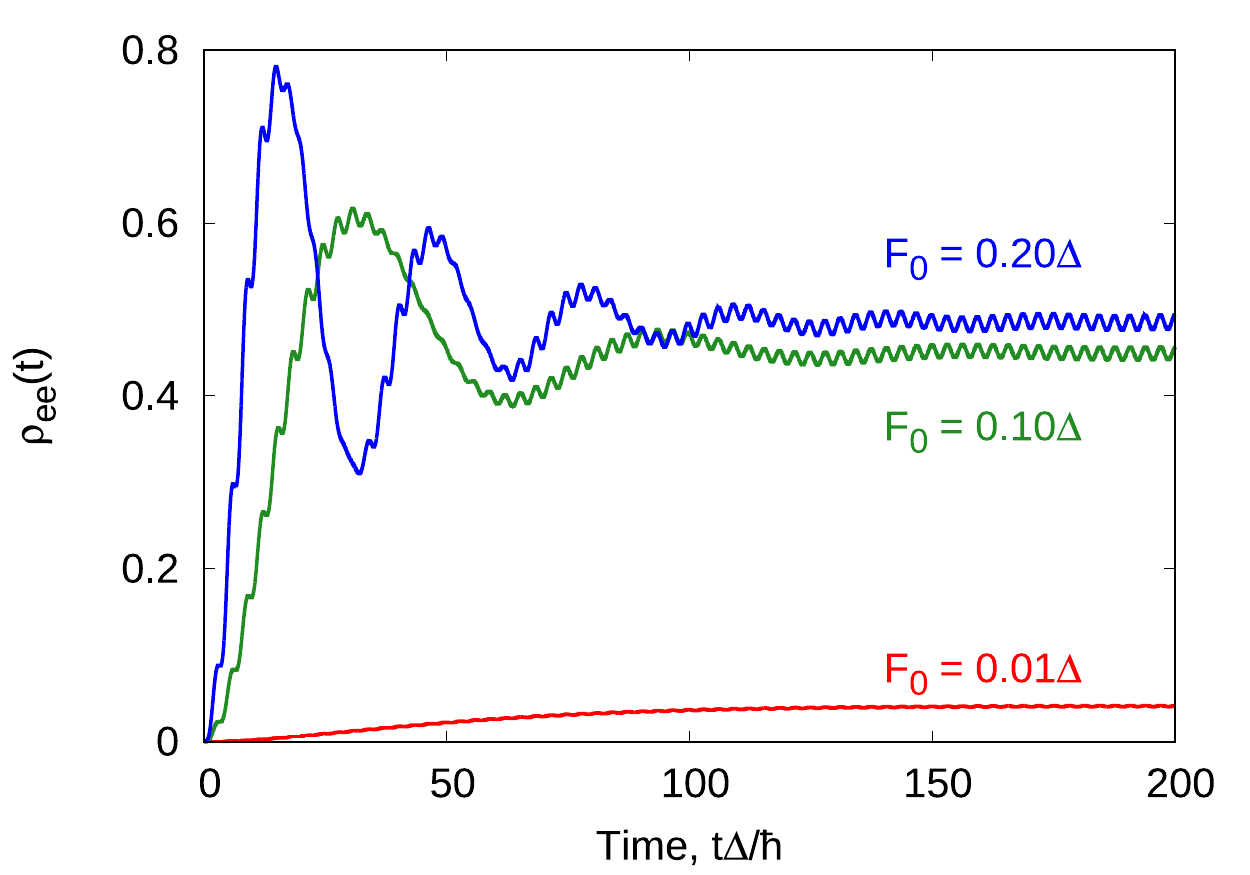}
\caption{\label{fig:pop_evolution}
Population dynamics of the driven two-level system with dissipation
under resonant driving~($\hbar \Omega=\Delta$) 
with different field strength $F_0$.
The relaxation times, $T_1$ and $T_2$, are set to $30\hbar/ \Delta$.
}
\end{figure}

Figure~\ref{fig:pop_evolution} shows the population of the excited state,
$\rho_{ee}(t)=\langle e|\rho(t)|e\rangle$, of the two-level system
as a function of time for different field strength, $F_0$.
Here, both the relaxation times, $T_1$ and $T_2$, are set to $30\hbar/\Delta$
in order to investigate the dynamics under a relatively weak-relaxation condition,
$T_1,T_2 \gg \hbar/\Delta$.
At the initial time~$(t=0)$, the excited population is zero as
the initial state is set to the ground state $|g\rangle$.
As seen from the figure, the excited population $\rho_{ee}(t)$ asymptotically reaches 
dynamics, which has the same time-periodicity as the external field $T$,
in the long propagation limit for each field strength.
In contrast, one sees that oscillatory features that have longer 
periodicity than $T$ are observed in the stronger field cases, 
and the periods of the oscillation depend on the field strength.
The period of the oscillatory feature is close to 
that of the Rabi oscillation, $T_R=2\pi/\Omega_R$,
where $\Omega_R$ is the Rabi frequency, $\Omega_R = F_0/\hbar$.
Thus, these oscillatory features can be understood as the Rabi oscillation
with damping due to the dissipation.
We note that, as seen from Fig.~\ref{fig:pop_evolution},
the timescale of approaching the steady state does not significantly depend on
the field strength, $F_0$. Because the lower bound of $T_1$ is determined by $T_2$
as $T_1 \ge T_2/2$, $T_1$ cannot be the relevant timescale independently.
Thus, the relevant timescale of approaching the steady
state is approximately determined by the decoherence time, $T_2$.

\begin{figure}[htbp]
  \includegraphics[width=0.95\columnwidth]{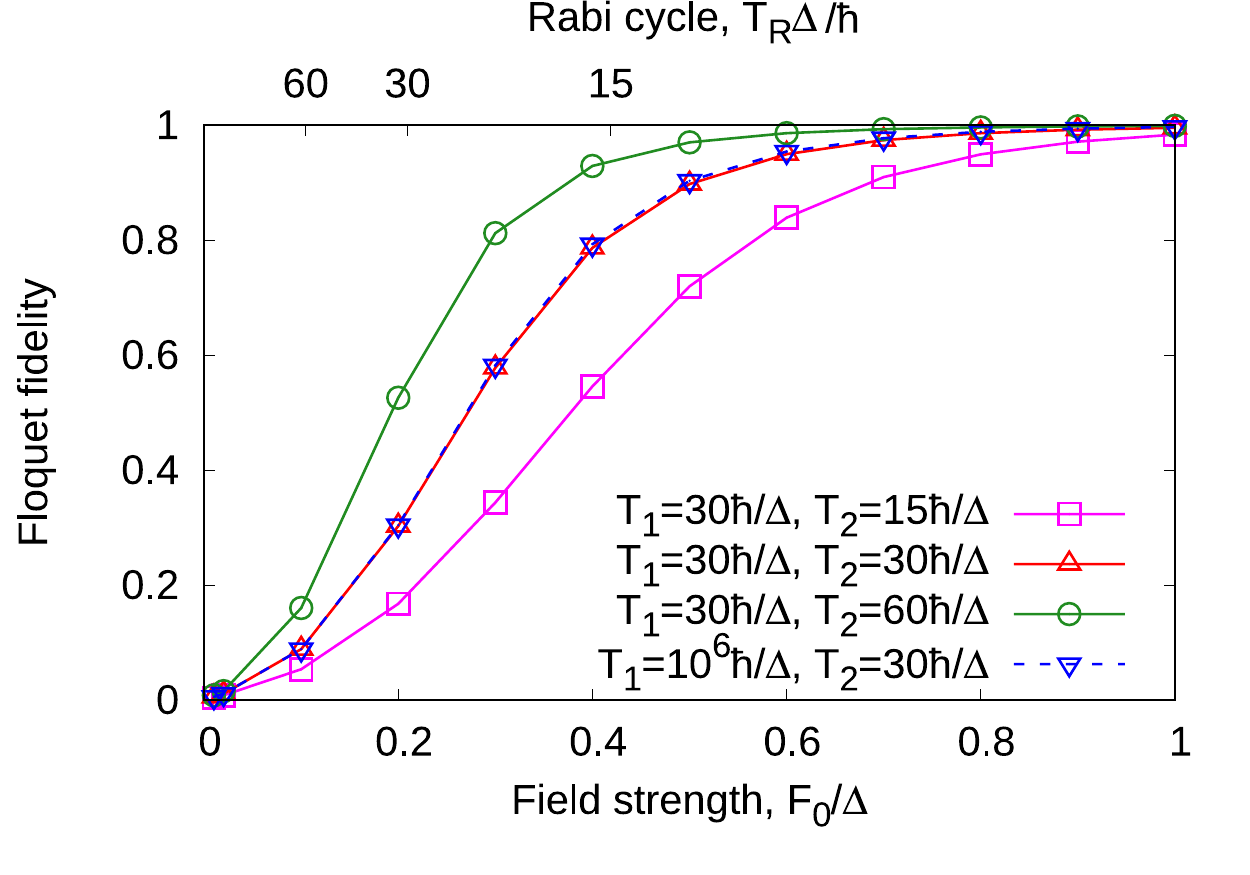}
\caption{\label{fig:floquet_fid_res}
Floquet fidelity of nonequilibrium steady states in the resonantly driving
condition, $\hbar \Omega = \Delta$.
The results with different relaxation conditions are shown as functions of 
applied field strength $F_0$. The secondary $x$-axis shows the corresponding
Rabi cycle, $T_R=2\pi/\Omega_R$ with the Rabi frequency $\Omega_R=F_0/\hbar$.
}
\end{figure}

Now we turn to studying the basic properties of the nonequilibrium steady state,
employing the Floquet fidelity, $S_F$.
Figure~\ref{fig:floquet_fid_res} shows the computed Floquet fidelity
of the nonequilibrium steady state as a function of driving field strength
$F_0$. The results with different relaxation conditions are shown in the figure.
The general feature is that, while the Floquet fidelity $S_F$ becomes zero in the weak
field limit, $S_F$ asymptotically reaches unity in the strong field limit.
This fact indicates that the Floquet states are significantly destructed
by the dissipation in the weak field regime.
In contrast, in the strong field regime, the contribution from the external driving
field overcomes the dissipation effect, and the Floquet states are stabilized.

In Fig.~\ref{fig:floquet_fid_res}, squares (purple), up-pointing triangles (red),
and circles (green) show the computed Floquet fidelity $S_F$ with
the same longitudinal relaxation time $T_1=30\hbar/\Delta$ 
but with different decoherence time $T_2$.
Comparing these results, one sees that the Floquet fidelity
becomes smaller when the decoherence time $T_2$ becomes shorter.
This fact indicates that the coherence plays an important role to form
the Floquet states, and the decoherence is a source of the destruction of the Floquet states.
In contrast, in Fig.~\ref{fig:floquet_fid_res}, up-pointing triangles (red)
and down-pointing triangles (blue) show the Floquet fidelity
with the same decoherence time $T_2=30\hbar/\Delta$ 
but different longitudinal relaxation time $T_1$. Despite the significant difference
of the longitudinal relaxation time $T_1$, the numerics provide
almost identical Floquet fidelities for all the investigated field strengths.
This fact clearly demonstrates that the population relaxation does not directly
affect the formation of Floquet states but it only affects the population of
the formed dressed states.
Therefore, the decoherence time $T_2$ is the only significant parameter for
the realization of the Floquet states, at least in the presently discussed Maxwell-Bloch equation.

Next we study the quasienergy spectrum of the driven open-quantum system, computed
by the stochastic trajectory approach, Eq.~(\ref{eq:pop-pes-open}),
employing a $\sin ^2$ envelope
for the probe field $f(t)$ with the total duration of $200\pi \hbar/\Delta$,
which is $100$ optical cycles of the pump field in the resonant condition,
$\hbar \Omega = \Delta$.

\begin{figure}[htbp]
  \includegraphics[width=0.95\columnwidth]{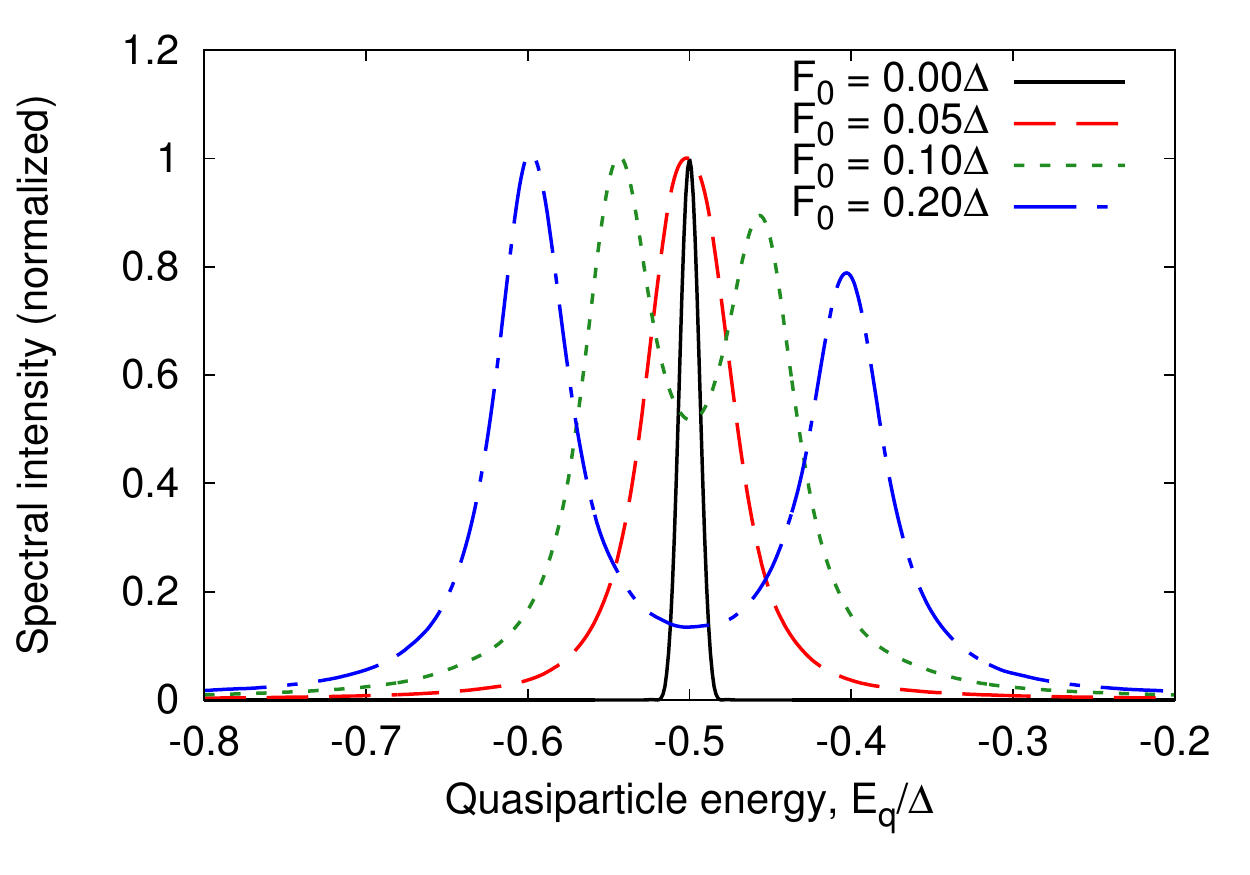}
\caption{\label{fig:pes_res}
Quasienergy spectrum computed by
Eq.~(\ref{eq:pop-pes-open}) under resonant driving ($\hbar \Omega=\Delta$).
The results for different driving field strength
are shown. Here, we set both the longitudinal and transverse relaxation times
to $T_1=T_2=30 \hbar/\Delta$.
Note the maximum value of the spectral density are normalized to one
for each field strength.
}
\end{figure}

Figure~\ref{fig:pes_res} shows the spectral density
as a function of quasienergy $E_q$, which is defined by
the difference of the photon-energy of the probe field $\hbar \omega$ and the energy
of the detector state $\epsilon$,
$E_q = \epsilon - \hbar \omega $.
Here, the relaxation times, $T_1$ and $T_2$, are set to $30\hbar/\Delta$.
The results computed with different field strength $F_0$ are compared in
Fig.~\ref{fig:pes_res}. The result without driving field~(black solid line) shows 
a peak at $-\Delta/2$, which is the single-particle energy of the ground state
$|g\rangle$. Because the quantum jump process in the stochastic approach
with the Lindblad operators, Eq.~(\ref{eq:L1-operator}) and
Eq.~(\ref{eq:L2-operator}), does not affect the ground state, $|g\rangle$,
the linewidth of the ground state spectrum is solely caused by the bandwidth
of the probe pulse.
When a driving field is applied, the quasienergy peak is broadened~(red-dashed line)
because the dissipative mechanism is activated by the photo-excitation.
Once the applied field strength becomes strong enough, the quasienergy peak is
split into two peaks, reflecting the well-known Rabi splitting
(see green-dotted and blue-dash-dot lines).
These results demonstrate that signatures of Floquet states are disturbed by
dissipation, and they can be evident only when the driving field strength is strong
enough to overcome the dissipation contribution.

\begin{figure}[ht]
  \includegraphics[width=0.95\columnwidth]{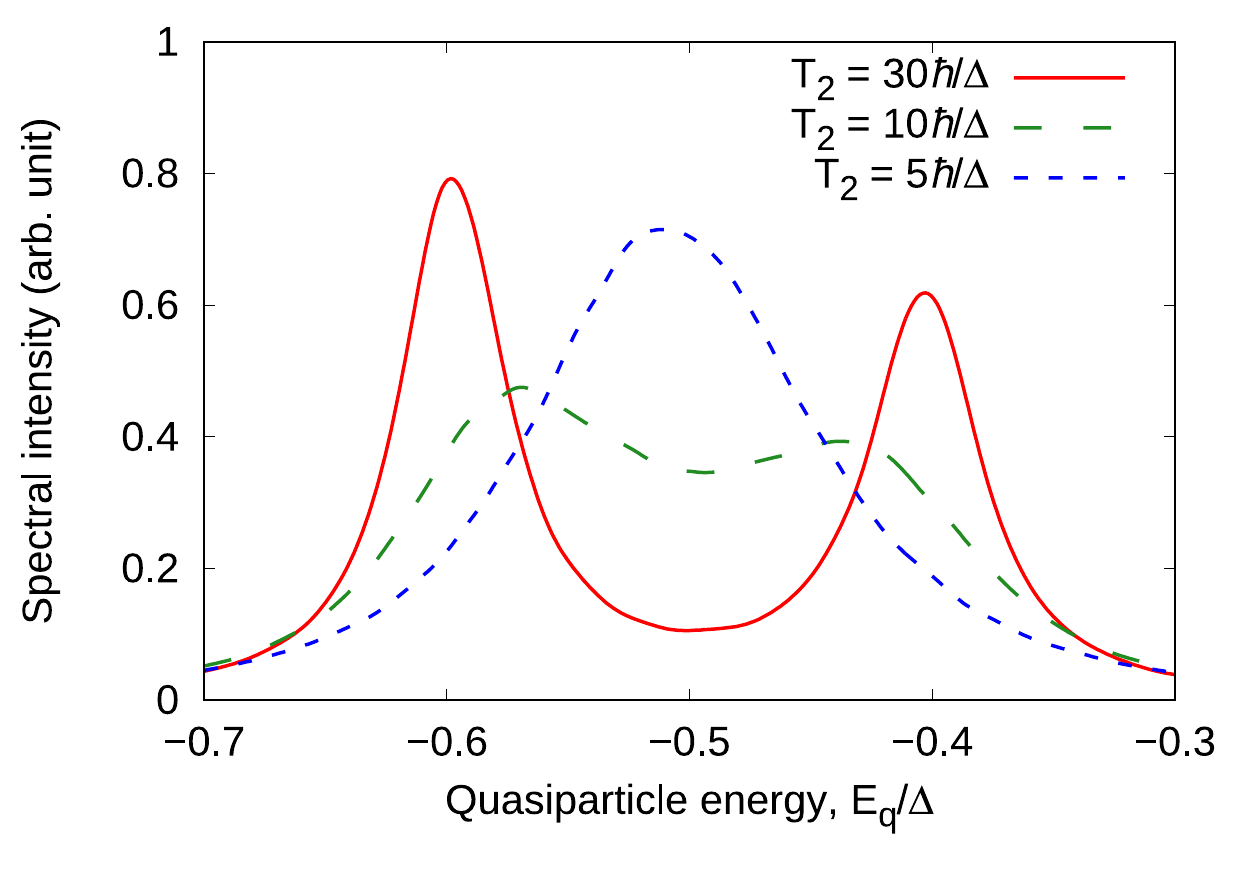}
\caption{\label{fig:pes_res_t2}
Quasienergy spectrum
computed by Eq.~(\ref{eq:pop-pes-open}) under resonant driving ($\hbar \Omega=\Delta$).
The results for different 
decoherence time $T_2$ are shown.
Here, we set the longitudinal relaxation time $T_1$ to $30 \hbar/\Delta$,
and the field strength $F_0$ to $0.2\Delta$.
}
\end{figure}

Let us now take a closer look at the role of the decoherence in the formation of Floquet states.
For this purpose, we compute the quasienergy spectrum
while also changing the decoherence time, $T_2$.
Figure~\ref{fig:pes_res_t2} shows the computed quasienergy spectra with
Eq.~(\ref{eq:pop-pes-open}).
In these calculations, the longitudinal relaxation time is fixed
to $T_1=30\hbar/\Delta$, and the field strength is fixed to
$F_0=0.2\Delta$. The period of the corresponding Rabi flopping is
$T_R=2\pi/F_0\approx 30\hbar/\Delta$.
The red-solid line in Fig.~\ref{fig:pes_res_t2} shows the result
with the decoherence time $T_2$ of $30\hbar/\Delta$, which is almost identical to
the period of the Rabi oscillation $T_R$, but it clearly shows the double peak structure
of the Rabi splitting, where the corresponding Rabi splitting energy is
$\hbar \Omega_R = F_0 = 0.2 \Delta$.
The green-dashed line shows the result with $T_2=10 \hbar/\Delta$,
which is almost one third of the Rabi cycle $T_R$.
The result clearly demonstrates that 
the key feature of Floquet states, namely Rabi-splitting, is still fairly visible
even though the decoherence time is substantially shorter than the Rabi cycle ($T_2<T_R$).
However, if the decoherence time $T_2$ is further halved and is set to 
$T_2=5\hbar/\Delta$, the double-peak structure disappears (blue dotted line).
This fact indicates that the coherence should survive for, at least, one third 
of the period of the Rabi oscillation in order to fairly observe the Rabi splitting.
Interestingly, by comparing the red-solid line and the blue-dotted line in Fig.~\ref{fig:pes_res_t2},
one can clearly see that the disappearance of the double-peak structure originates from
not only the line-broadening but also the collapse of the gap.
This fact further implies that the formation of the Floquet states are
significantly disturbed due to loss of coherence.

\begin{figure}[ht]
  \includegraphics[width=0.95\columnwidth]{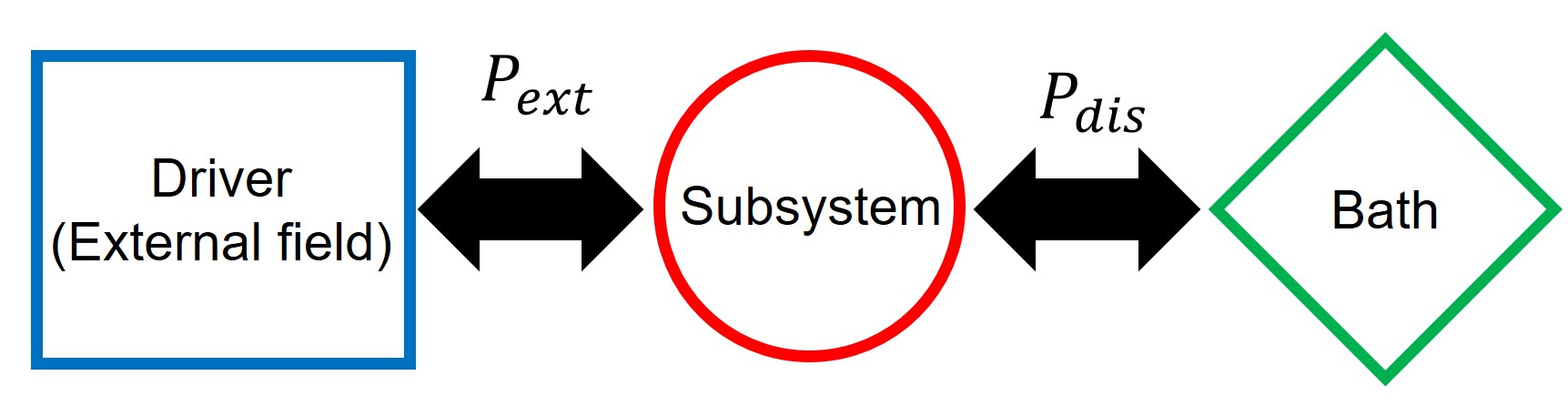}
\caption{\label{fig:energy_flow}
Schematic diagram of the energy flow among the external driver (external field),
the subsystem, and the bath.
Two kinds of energy flow exist: One is the flow from the external driver to the subsystem
$P_{ext}$, and the other is that from the bath to the subsystem $P_{dis}$.
}
\end{figure}

Next, we explore the role of the dissipation in the nonequilibrium steady state
based on an analysis of the microscopic energy flow.
Figure~\ref{fig:energy_flow} schematically shows the energy flow among
the external driver (external field), the subsystem, and the bath.
As seen from the figure, we consider two kinds of energy flow, $P_{ext}(t)$ and
$P_{dis}(t)$:
$P_{ext}(t)$ is the energy flow from the external field to the subsystem,
and $P_{dis}(t)$ is that from the environment (dissipation) to the subsystem.
The energy flow from the external field to the subsystem $P_{ext}$ can be evaluated
with the Joule heating
(see Appendix \ref{appendix:joule-heating} for details)
as
\be
P_{ext}(t) &=& \mathrm{Tr}\left \{ \frac{1}{i\hbar} \left[ 
\frac{\Delta}{2}\sigma_z ,H(t)\right ] \rho(t) 
\right\} \nonumber \\
&=& \frac{\Delta F_0}{\hbar} \sin(\Omega t) \mathrm{Tr}\left \{ \sigma_y \rho(t)
\right \}.
\label{eq:joule-heating}
\ee
Because of the total energy conservation law, the energy change of the subsystem
has to be identical to the sum of the energy transfer as
\be
\frac{d}{dt}E_s(t) = P_{ext}(t) + P_{dis}(t),
\label{eq:tota-energy-change}
\ee
where $E_s(t)$ is the energy of the subsystem,
$E_s(t)=\mathrm{Tr}\left \{ \Delta \sigma_z\rho(t)\right\}/2$.
Based on this fact, we redefine $P_{dis}(t)$ as
\be
P_{dis}(t) \equiv \frac{d}{dt}E_s(t) - P_{ext}(t).
\ee

\begin{figure}[htbp]
  \includegraphics[width=0.95\columnwidth]{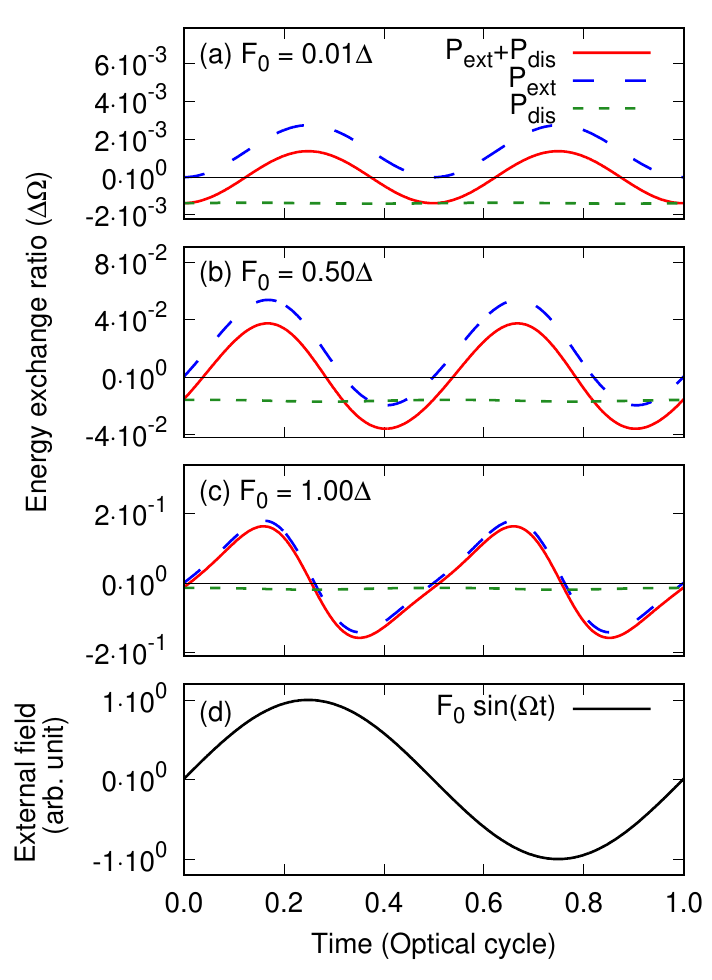}
\caption{\label{fig:energy_exchange}
Energy flow in the nonequilibrium steady state among the subsystem,
the external driver and the environment under resonant driving~($\hbar \Omega=\Delta$)
as a function of time.
The results for different field strength are shown:
(a)~$F_0=0.01\Delta$, (b)~$F_0=0.50\Delta$, and (c)~$F_0=1.00\Delta$.
}
\end{figure}

Figure~\ref{fig:energy_exchange} shows the energy flow in the nonequilibrium
steady state as a function of time for different field strength $F_0$.
Here, we set both the relaxation times, $T_1$ and $T_2$, to $30 \hbar/\Delta$.
As seen from Fig.~\ref{fig:energy_exchange}~(a),
the energy flow from the light to the subsystem $P_{ext}(t)$ (green-dashed line) 
is always positive while that from the environment to the subsystem
$P_{dis}(t)$ (blue-dotted line) is always negative.
Therefore, the transferred energy from the external field does not return to
the external driver but it is completely dissipated by the environment
in the weak field regime. Hence, the energy exchange between the subsystem
and the external driver is significantly disturbed by the dissipation, and the formation
of Floquet states is prevented.
As shown in Fig.~\ref{fig:energy_exchange}~(b),
once the field strength becomes substantially strong~($F_0=0.5\Delta$),
the energy flow $P_{ext}(t)$ shows a negative value around a certain time.
The corresponding Floquet fidelity $S_F$ for this field strength is about $0.9$
(see Fig.~\ref{fig:floquet_fid_res}).
This fact indicates that the transferred energy from the external driver to the subsystem
is not completely dissipated to the environment, but a part of the transferred energy
is returned to the external driver. Thus, the energy exchange between the subsystem
and the external driver becomes possible, and the corresponding Floquet states are fairly
formed.
As shown in Fig.~\ref{fig:energy_exchange}~(c), once the field strength becomes
very strong~($F_0=\Delta$), the energy flow $P_{ext}(t)$ becomes dominant, compared with
the dissipation $P_{dis}(t)$, and almost all of the transferred energy from
the external driver to the subsystem returns back to the driver.
As a result, the corresponding Floquet fidelity $S_F$ becomes almost unity
(see Fig.~\ref{fig:floquet_fid_res}), and the Floquet states are almost perfectly realized.

To comprehensively study the role of $T_1$ and $T_2$, we further repeated the energy
flow analysis with different relaxation conditions
(see Appendix~\ref{appendix:energy-exchange-t1-t2} for details).
As a result, we found that the qualitative behavior of the energy flow
does not depend on $T_1$ while it can be affected by $T_2$. This fact further indicates
that the longitudinal relaxation characterized by $T_1$ does not disturb the energy
exchange between the subsystem and the external driver, and it does not disturb
the formation of Floquet states. In contrast, the decoherence characterized by $T_2$
can disturb the energy exchange and the formation of Floquet states.

Based on the above analysis, the energy exchange between the system and the driving field
is expected to play an important role to realize Floquet state as well as
the photo-dressed states.
This results may further indicate a possibility to stabilize Floquet states
by tuning the energy exchange with additional controlling fields such as a secondary
laser field. A possibility of stabilization of Floquet states with multi-color laser fields
will be investigated in future work based on these findings.

At the end of this subsection, we investigate the nonequilibrium steady state 
under the significant decoherence, where the decoherence time is substantially 
shorter than the cycle of the external driving.
Thus, the coherence does not survive even for the single period
of the driving field. Under such significant decoherence, can Floquet states
be still realized? To address this question, we investigate the nonequilibrium steady
state by setting $T_1$ to $30\hbar/\Delta$ and $T_2$ to $T_{cycle}/2$, which
is the half cycle of the external driving.

\begin{figure}[htbp]
  \includegraphics[width=0.95\columnwidth]{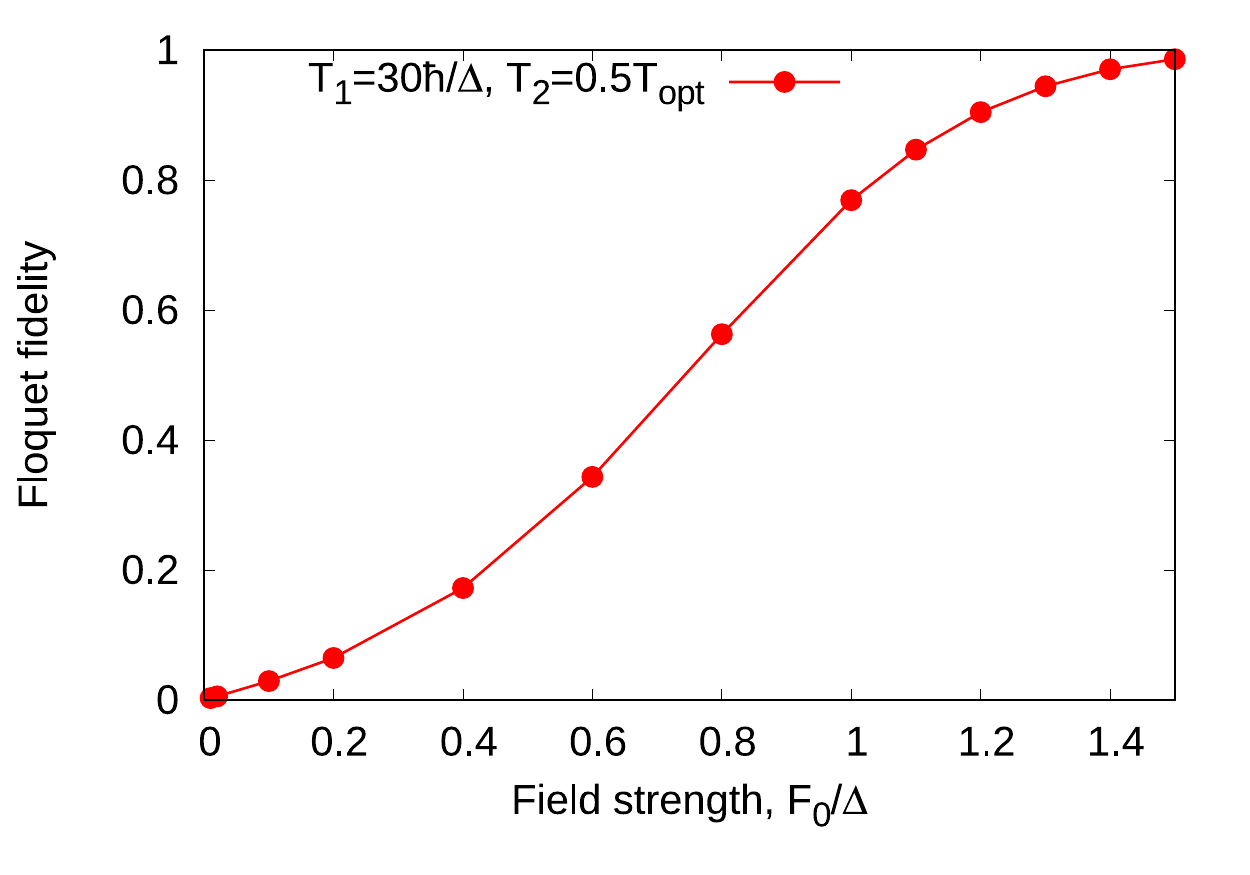}
\caption{\label{fig:floquet_fid_res_t2_pi}
Floquet fidelity as a function of the field strength $F_0$
in the resonantly driving condition, $\hbar \Omega = \Delta$.
Here, we set the driving frequency $\Omega$ to $\Delta/\hbar$,
the longitudinal relaxation time $T_1$ to $30\hbar/\Delta$,
the decoherence time $T_2$ to the half cycle of the driving $T_{cycle}/2$.
}
\end{figure}

Figure~\ref{fig:floquet_fid_res_t2_pi} shows the computed Floquet fidelity
as a function of the driving field strength $F_0$.
In the weak field limit, the Floquet fidelity becomes zero, indicating that
the Floquet states is significantly disturbed
by the decoherence. In contrast, the Floquet fidelity asymptotically
reaches to one in the strong field regime, indicating that the decoherence
effect is overcome by the strong driving, and the Floquet states are stabilized.
This result clearly demonstrates that Floquet states can be realized
with a sufficiently strong driving field even under the influence of 
significant decoherence, where the coherence is lost before 
the single-cycle of the external driving.

\subsection{Off-resonance \label{subsec:off-resonance}}

Here, we investigate the nonequilibrium steady state 
in an off-resonant condition. 
For this purse, we set the driving frequency $\Omega$ of the field to one third of the gap of the system, $\Delta/3\hbar$.
This is nothing else than the three-photon resonance condition.
Note that the three-photon absorption process is the lowest order nonlinear
photo-excitation in the present model because
the even-photon absorption processes, including the two-photon absorption, are 
forbidden by the symmetry of the Hamiltonian.
In this subsection, we further set both the relaxation times, $T_1$ and $T_2$,
to $30\hbar/\Delta$.

\begin{figure}[ht]
  \includegraphics[width=0.95\columnwidth]{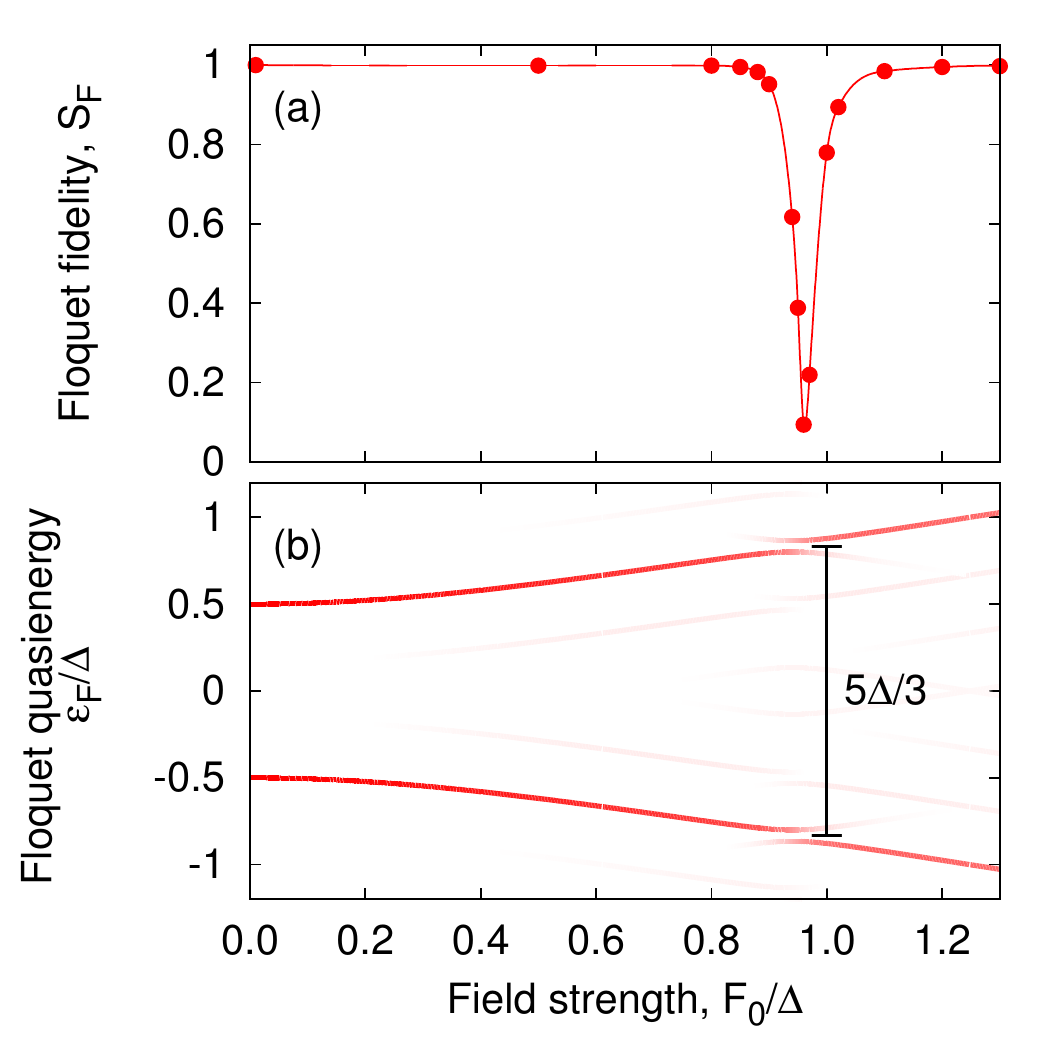}
\caption{\label{fig:floquet_fid_3ph}
(a)~Floquet fidelity as a function of the field strength $F_0$
in the off-resonant condition, $\hbar \Omega = \Delta/3$.
Here, we set both the relaxation times, $T_1$ and $T_2$, to $30\hbar/\Delta$.
(b)~the Floquet quasienergy spectrum as a function of the field strength $F_0$
in the off-resonant condition.
}
\end{figure}

Figure~\ref{fig:floquet_fid_3ph}~(a) shows the computed Floquet fidelity
$S_F$ as a function of field strength $F_0$.
In the weak field regime, the Floquet fidelity is close to one,
indicating that the Floquet states are almost perfectly realized.
This behavior is qualitatively different from that in the resonant condition
(see Fig.~\ref{fig:floquet_fid_res}):
in the resonant condition, the Floquet fidelity is almost zero in the weak field regime.
The qualitative difference of the two conditions can be explained by the photo-induced
population transfer \cite{PhysRevB.99.214302,2019arXiv190512981S}:
While the dissipation mechanism is activated in the resonant condition
because the excited state is populated by the resonant excitation,
it is not activated in the off-resonant condition because
the population transfer cannot occur due to the energy gap.
Once the field strength becomes substantially strong, the Floquet fidelity
becomes small, indicating that the Floquet states are disturbed by the dissipation.
Then, when the field strength becomes even stronger, the Floquet fidelity
approaches to one again.
To elucidate the mechanism of the temporal reduction of the Floquet fidelity
in Fig.~\ref{fig:floquet_fid_3ph}~(a),
we evaluate Floquet quasienergies based on the Fourier decomposition of the Floquet states,
\be
|\psi_{F,a}(t)\rangle=\sum_m e^{-i\frac{\epsilon_{a,m}}{\hbar}t} |u_{a,m}\rangle, 
\ee
where $\epsilon_{a,m}$ is the replicated Floquet-quasienergy defined as
$\epsilon_{a,m}=\epsilon_{a}+m\hbar\Omega$,
and $|u_{a,m}\rangle$ is the corresponding Fourier component.

Figure~\ref{fig:floquet_fid_3ph}~(b) shows the computed Floquet quasienergy
$\epsilon_{a,m}$ as a function of the applied field strength, $F_0$.
The false color shows the norm of the corresponding state,
$\langle u_{a,m}|u_{a,m}\rangle$.
In the weak field limit, the states have the bare gap of $\Delta$.
As the field strength increases, the gap becomes larger due to the dynamical Stark effect.
When the field strength $F_0$ is close to $\Delta$, the gap between the dominantly populated
states reaches $5\Delta/3$, which is identical to five times
the photon-energy of the applied field, $\hbar \Omega=\Delta/3$.
Therefore, the five-photon absorption process is expected to occur
around this field strength.
Indeed, the Floquet quasienergy spectrum in Fig.~\ref{fig:floquet_fid_3ph}~(b)
clearly shows the energy splitting around this field strength.
Evidently, the Floquet fidelity $S_F$ is sharply reduced around this five-photon absorption
regime, comparing Figs.~\ref{fig:floquet_fid_3ph}~(a) and (b).
Therefore, the destruction of the Floquet states can be understood as the activation
of the dissipative mechanism through multi-photon processes.
Importantly, the three-photon absorption process, which is the lowest possible multi-photon
absorption process does not have a substantial impact on the activation of the dissipation
because it is significantly suppressed by the band-gap renormalization due to
the dynamical Stark effect.

The above finding indicates that the population transfer (heating effect) has
a significant impact in the disappearance of
the Floquet states even in the off-resonant condition.

\section{Summary \label{sec:summary}}
In this work, we investigated some basic properties of nonequilibrium steady states
driven by periodic driving fields under the influence of dissipation.
We employed the Maxwell-Bloch equation~\cite{1072212,meier2007coherent}
and equivalent formulations in order to evaluate
the Floquet fidelity $S_F$ and the quasienergy spectrum of the nonequilibrium steady state.

First, we investigated the properties of the nonequilibrium steady state
in the resonant driving condition. In the weak field strength limit,
the Floqeut fidelity approaches to zero. This fact indicates that
the Floquet states are significantly destructed by the system-environment interaction
that is triggered by the photoexcited population $\rho_{ee}(t)=\langle e|\rho(t)|e\rangle$.
When the field strength becomes substantial, the Floquet fidelity monotonically increases
and asymptotically approaches to one, reflecting that the nonequilibrium steady states are 
perfectly described by the Floquet states.
This behavior can be understood in terms of the competition of the driving field contribution
and the dissipation contribution. In the weak field regime, the driving field contribution
is overcome by the dissipation contribution. As a result, the Floquet states 
are significantly disturbed, and the Floquet fidelity $S_F$ becomes small.
Once the field strength becomes strong enough, the driving field contribution
becomes dominant, compared with the dissipation contribution. As a result, the Floquet
states are stabilized, and the Floqeut fidelity approaches to one.

To elucidate the detailed roles of the dissipation, we evaluated the Floquet fidelity by
varying the relaxation times, $T_1$ and $T_2$.
As a result, we found that the longitudinal relaxation time $T_1$ does not have
a direct impact on the formation of Floquet states while 
the transverse relaxation time (decoherence time) $T_2$ has a significant impact on 
the formation and the destruction of Floquet states.
These results indicate that the coherence plays an important role in the formation
of Floquet states and it has to survive for a relevant timescale to realize Floquet states.

Then, employing the stochastic wavefunction approach, we investigate 
the quasienergy structure of the nonequilibrium steady state in 
the resonant condition. Consistently with the above Floquet fidelity analysis,
the quasienergy spectrum shows the double peak structure
as a signature of the Rabi-splitting once the applied field strength becomes
strong enough. To elucidate the role of the decoherence in the formation of 
the Floquet features, we computed the energy spectrum 
by varying the decoherence time $T_2$ (see Fig.~\ref{fig:pes_res_t2}).
As a result, we found that the decoherence destructs the feature of the Floquet states
in the energy spectrum by causing the collapse of the gap of the Rabi splitting.
This fact clearly demonstrates that the decoherence does not only hide the Floquet features
in the spectrum due to the line-broadening but also destructs the Floquet states themselves.
Therefore, the suppression of the decoherence effects is expected to be very important
to optically control material functionalities via the Floquet engineering.

Next, we studied the nonequilibrium steady state under the influence of the significant
decoherence in order to address the following question: Can Floquet states be
formed even if the coherence is annihilated before the optical cycle?
As a result of the analysis, we demonstrated that the Floquet states
can indeed be formed even under the significant decoherence once the field strength becomes
strong enough (see Fig.~\ref{fig:floquet_fid_res_t2_pi}).
This fact indicates that the period of driving fields
is not a relevant timescale for the formation of Floquet states.

Finally, we investigated the Floquet fidelity in the off-resonant condition,
where the photon-energy of the driving field is set to the one third of the gap,
$\hbar \Omega=\Delta/3$. In the off-resonant condition,
the Floquet fidelity becomes almost one 
in the weak field regime in contrast to the resonant condition.
This result indicates that the Floquet states are well formed
in the off-resonant weak field regime because the photo-excitation is forbidden
by the gap and the dissipation contribution is not activated.
Furthermore, we found that the Floquet fidelity can be substantially reduced
once the multi-photon excitation becomes relevant because the photo-excitation
further triggers the dissipation mechanism and the Floquet states are disturbed by
the system-environment interaction.

The above findings clearly demonstrate that heating effects and/or  photo-excitation
effects can significantly affect the formation of Floquet states because
the excess energy of excited systems can be dissipated to its environment through
the system-environment interaction, which further destructs
the coherence of the field driven dynamics.
Therefore, one can expect that the Floquet states may be stabilized
by reducing the effective energy dissipation to the environment
with additional external driving fields. For example, one may realize the stabilized
Floquet states with multi-color laser fields; one color mainly drives the Floquet states,
and the other colors stabilize them by renormalizing the energy dissipation.
This is also known as  optical-control of coherence through the control of dissipation,
and it may further introduce additional degree of freedoms in the Floquet engineering
and the optical-control itself.

In this work, we employed the simplest Markovian master equation to clarify
the primary role of dissipation in driven quantum systems.
Therefore, a role of \textit{memory effects} in driven quantum systems,
especially in the context of the formation of dressed states, has not been explored yet.
Although theoretical treatment of such memory effects with non-Markovian master
equations is much more difficult than the simple Markovian treatment,
the role of the memory effects has to be clarified towards the optical-control
of material functionalities and phases of matter because rich physical properties
may be realized in driven quantum systems relaying on complex memory effects.
Work along these lines with non-Markovian quantum master equations is already under way.

\begin{acknowledgements}
We acknowledge fruitful discussions with M.~A.~Sentef and P.~Tang.
This work was supported by the European Research Council (ERC-2015-AdG694097)
and JST-CREST under Grant No. JP-MJCR16N5.
The Flatiron Institute is a division of the Simons Foundation.
S.A.S. gratefully acknowledges the fellowship from the Alexander von Humboldt Foundation.
A.R. acknowledges support from the Cluster of Excellence 'Advanced Imaging of Matter' (AIM).
\end{acknowledgements}

\appendix

\section{Energy transfer from external fields to
quantum systems
\label{appendix:joule-heating}}

Here, we revisit the energy transfer from an external field to
the quantum two-level system.
To purely evaluate the energy exchange between the external field and the quantum system,
we disregard the dissipation and assume that the dynamical system is
described by the following quantum Liouville equation
\be
\frac{d}{dt}\rho(t) = \frac{\left [ H(t),\rho(t) \right ]}{i\hbar},
\label{eq;appendix-liouville}
\ee
with the $2\times 2$ Hamiltonian matrix
\be
H(t) = \frac{\Delta}{2}\sigma_z + F_0 \sin (\Omega t) \sigma_x.
\ee

The energy of the quantum system is defined with the unperturbed Hamiltonian
$\Delta \sigma_z/2$ as
\be
\tilde E_s(t) = \mathrm{Tr}\left \{\frac{\Delta}{2}\sigma_z \rho(t)
\right \}.
\ee
Thus the energy change of the subsystem by the external field is evaluated as
\be
P_{ext}(t) &\equiv& \frac{d}{dt} \tilde E_s(t) \nonumber \\
&=&\mathrm{Tr}\left \{ \frac{1}{i\hbar} \left[ 
\frac{\Delta}{2}\sigma_z ,H(t)\right ] \rho(t) 
\right\} \nonumber \\
&=& \frac{\Delta F_0}{\hbar} \sin(\Omega t) \mathrm{Tr}\left \{ \sigma_y \rho(t)
\right \}.
\ee
This is nothing but the energy gain of the subsystem purely from the external field,
and it is introduced as $P_{ext}(t)$ in Eq.~(\ref{eq:joule-heating}).

In the main text, we further define the energy flow from the environment
as the difference between the total energy change $dE_s(t)/dt$ and
the pure external-field contribution $P_{ext}(t)$ in Eq.~(\ref{eq:tota-energy-change}).

\section{Energy exchange analysis with several relaxation conditions
\label{appendix:energy-exchange-t1-t2}}

For a comprehensive study, we repeat the energy flow analysis shown in
Fig.~\ref{fig:energy_exchange} with different relaxation conditions.
Note that, in the analysis of Fig.~\ref{fig:energy_exchange}, the relaxation times,
$T_1$ and $T_2$, are set to $30\hbar/\Delta$.

\begin{figure}[htbp]
  \includegraphics[width=0.95\columnwidth]{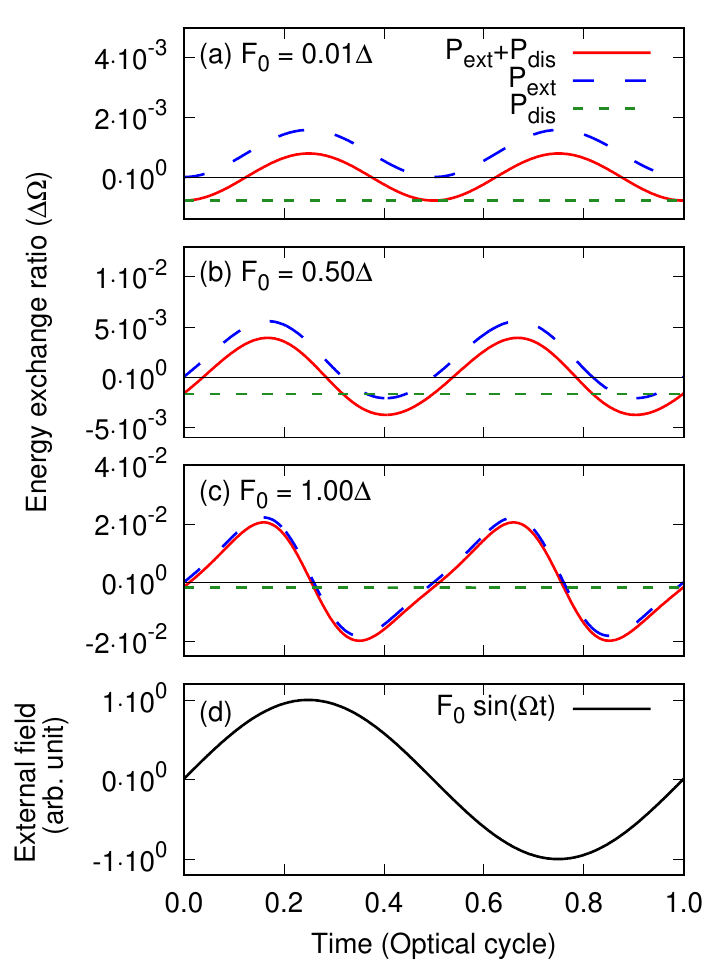}
\caption{\label{fig:energy_exchange_t1_mod}
Energy flow in the nonequilibrium steady state among the subsystem,
the external driver and the environment under resonant driving~($\hbar \Omega=\Delta$)
as a function of time. Here, $T_1$ is set to $300\hbar/\Delta$ and
$T_2$ is set to $30\hbar/\Delta$.
The results for different field strength are shown:
(a)~$F_0=0.01\Delta$, (b)~$F_0=0.50\Delta$, and (c)~$F_0=1.00\Delta$.
}
\end{figure}

First, we investigate the effect of the longitudinal relaxation time $T_1$ in
the energy flow. For this purpose, we set $T_1$ to $300\hbar/\Delta$, which 
is ten times larger than the original analysis in Fig.~\ref{fig:energy_exchange},
while fixing $T_2$ to the original value, $ 30\hbar/\Delta$.
Figure~\ref{fig:energy_exchange_t1_mod} shows the computed energy flow with different field
strength. Comparing Fig.~\ref{fig:energy_exchange_t1_mod} with
Fig.~\ref{fig:energy_exchange}, one sees that the qualitative behavior of the energy flow
in the two relaxation conditions does not change despite the significant difference
of the longitudinal relaxation time, $T_1$.
Therefore $T_1$ does not affect the energy exchange between
the subsytem and the external driver.

\begin{figure}[htbp]
  \includegraphics[width=0.95\columnwidth]{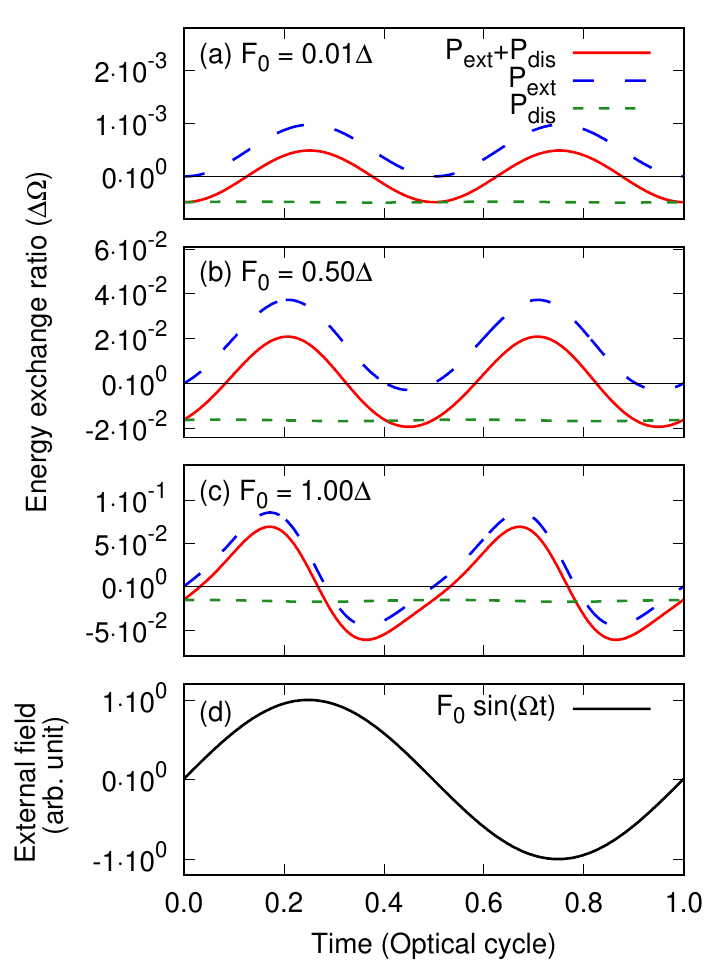}
\caption{\label{fig:energy_exchange_t2_mod}
Energy flow in the nonequilibrium steady state among the subsystem,
the external driver and the environment under resonant driving~($\hbar \Omega=\Delta$)
as a function of time. Here, $T_1$ is set to $30\hbar/\Delta$ and
$T_2$ is set to $10\hbar/\Delta$.
The results for different field strength are shown:
(a)~$F_0=0.01\Delta$, (b)~$F_0=0.50\Delta$, and (c)~$F_0=1.00\Delta$.
}
\end{figure}

Next, we investigate the effect of the transverse relaxation time $T_2$ in
the energy flow. For this purpose, we set $T_2$ to $10\hbar/\Delta$, which
is three times shorter than the original analysis in Fig.~\ref{fig:energy_exchange},
while fixing $T_1$ to the original value, $ 30\hbar/\Delta$.
Figure~\ref{fig:energy_exchange_t2_mod} shows the computed energy flow with different field
strength. In the weak field regime ($F_0=0.01\Delta$),
Fig.~\ref{fig:energy_exchange_t2_mod}~(a) and Fig.~\ref{fig:energy_exchange}~(a) 
do not show the qualitative difference because all the transferred energy to the system
is dissipated and no energy returns to the external driver.
In contrast, by comparing Fig.~\ref{fig:energy_exchange_t2_mod}~(b) and
Fig.~\ref{fig:energy_exchange}~(b), one can clearly see that the larger ratio of
the transferred energy is dissipated in the case of the stronger decoherence
($T_2=10\hbar/\Delta$) compared with the weaker decoherence ($T_2=30\hbar/\Delta$).
Therefore $T_2$ can directly affect the energy exchange between the subsystem and
the external driver.

Based on the above findings, we conclude that the longitudinal relaxation with $T_1$
 does not affect the energy exchange between the subsystem and the external driver while
the transverse relaxation with $T_2$ can significantly affect the energy exchange.
This conclusion may be counterintuitive because the longitudinal relaxation with $T_1$
directly links the energy dissipation while the transverse relaxation with $T_2$
does not change the subsystem energy when the subsystem is undriven ($F_0=0$).
The apparent inconsistency can be explained by the efficiency of the energy
return to the external driver with coherent driving: If the subsystem keeps
the perfect coherence ($T_2=\infty$), the subsystem shows the Rabi flopping,
realizing the perfect energy exchange as all the transferred energy to the subsystem
returns to the external driver. However, once the coherent dynamics is disturbed by
the decoherence, the perfect Rabi flopping is destroyed and all the transferred
energy cannot return to the driver anymore. In this regard, the efficiency of
the energy return is affected by the decoherence.
Since the subsystem is connected to the bath, the unreturned energy is simply dissipated
to the bath. This scenario can explain the apparent inconsistency of the energy dissipation
and the relaxation times, $T_1$ and $T_2$, further indicating the significance of
the preservation of coherence in the driven dynamics to realize Floquet states.

\bibliography{ref}

\end{document}